\documentclass{amsart}

\usepackage{
	algorithm,algpseudocode,
	amsmath,amsthm,amssymb,amsfonts,amscd,amsxtra,
	caption,color,courier,
	gensymb,hhline,kotex,lipsum, longtable,
	mathrsfs,mathtools,mdframed,multirow,
	soul,subcaption,stackrel,textcomp,tikz,
	verbatim,xcolor}
\usepackage[foot]{amsaddr}

\mathtoolsset{showonlyrefs}

\theoremstyle{plain}
\newtheorem{thm}{Theorem}

\newtheorem*{thm*}{thm}
\newtheorem*{lem*}{Lemma}
\newtheorem*{prop*}{Proposition}
\newtheorem*{cor*}{Corollary}
\newtheorem*{conj*}{Conjecture}

\theoremstyle{definition}

\newtheorem{exmp}[thm]{Example}

\newtheorem{rem}[thm]{Remark}

\title[A degree reduction for the graph coloring problem]{A degree reduction method for an efficient QUBO formulation for the graph coloring problem}

\author[N. Hong]{Namho Hong$^{1}$}
\author[H. Jung]{Hyunwoo Jung$^{1}$}
\author[H. Kang]{Hyosang Kang$^{2,\ast}$}
\author[H. Lim]{Hyunjin Lim$^{1}$}
\author[C. Seol]{Chaehwan Seol$^{3}$}
\author[S. Um]{Seokhyun Um$^{1}$}

\address{$^{1}$Taegu Science High School (TSHS), Daegu 42110, South Korea.}
\address{$^{2}$Daegu Gyeongbuk Institute of Science and Technology (DGIST), Daegu 42988, South Korea.}
\address{$^{3}$Gwangju Science Academy (GSA), Gwangju 61005, South Korea.}
\address{$^{\ast}$ Corresponding author}
\email{hyosang@dgist.ac.kr}

\thanks{This work was supported by TSHS R\&E grant at Daegu Gyeongbuk Institute of Science and Technology (DGIST). (grant number 2022030163)}

\begin{document}

\begin{abstract}
We introduce a new degree reduction method for homogeneous symmetric polynomials on binary variables that generalizes the conventional degree reduction methods on monomials introduced by Freedman and Ishikawa.
We also design an degree reduction algorithm for general polynomials on binary variables,
simulated on the graph coloring problem for random graphs,
and compared the results with the conventional methods.
The simulated results show that our new method produces reduced quadratic polynomials that contains less variables than the reduced quadratic polynomials produced by the conventional methods.
\end{abstract}



\maketitle



\section{Introduction}\label{sec:intro}

A graph is a 1-dimensional object that consists of vertices and edges. 
Two vertices are called adjacent if they are connected by an edge, and a graph is called simple if there is neither a loop nor multiple edges that joins adjacent vertices.
A vertex coloring is an assignment of ``colors" to vertices in a simple graph in a way that no two adjacent vertices are assigned by the same color.
The chromatic number of a simple graph is the minimum number of colors that a vertex coloring of the graph exists.
The graph coloring problem is a problem that concerns about finding the chromatic number and the existence of a  vertex colorings with a certain number of colors. 
It applies to a range problems such as scheduling \cite{Le79}, register allocation in compilers \cite{Ch82}, and frequency assignment in wireless communications \cite{Ba06}. 
The graph coloring problem is NP-hard, meaning that there is no known polynomial-time algorithm that solves the problem unless P=NP \cite{Ga76, Ga90}.

For graphs with large numbers of vertices and edges, finding a solution to the graph coloring problem may require lots of computational time and resources.
To circumvent this, one can use a quantum annealing machine, such as D-Wave's quantum annealer, which relies on the quantum tunneling effect \cite{Ap89, Ka98}.
A quantum annealing system can solve optimization problems whose objective functions are formulated as polynomials of binary variables, when there is no constraining equations.
The number of qubits in a quantum annealing machine limits the number of binary variables that the objective function can have.
Although there are few algorithms, such as minor-embedding, that breaks down a large problem into a set of smaller problems that a quantum annealing machine can handle \cite{W20}, it is highly desired to formulate the original objective function with as few variables as possible.

An optimization problem that the objective function is given as a polynomial of degree two with binary variables and no constraining equation is called QUBO (quadratic unconstrained binary optimization) \cite{Si20, Ta20}.
D-Wave's quantum annealer can find a solution of QUBO, i.e. it can find the global minimum of a quadratic polynomial of binary variables.
However, many of discrete optimization problems, such as K-SAT and graph coloring problems, are formulated with the objective functions as polynomials of the degree higher than two \cite{Si13}.
Thus one needs to reformulate the objective functions as quadratic polynomials by using degree reduction methods \cite{Sa18}.

Let $p$ be a polynomial that consists of $n$ binary variables $x_0,\cdots,x_{n-1}$.
A \textbf{degree reduction method} is a method of finding a polynomial $q$ of the degree less than the degree of $p$ that consists of binary variables $x_0,\cdots,x_{n-1}$ and extra variables $w_0,\cdots,w_{d-1}$, called \textbf{auxiliary variables}, that attains the same values of $p$ for any given values for $x_0,\cdots,x_{n-1}$.
More precisely, we will say that the polynomial $q(x_0,\cdots,x_{n-1},w_0,\cdots,w_{d-1})$ is \textbf{reduced} from $p(x_0,\cdots,x_{n-1})$ if $\deg q<\deg p$ and
\begin{equation}
	p(x_0,\cdots,x_{n-1}) = \min_{\substack{w_i=0,1\\i=0,\ldots,d-1}}q(x_0,\cdots,x_{n-1},w_0,\cdots,w_{d-1})
\end{equation}
for any binary values for $x_0, \cdots,x_{n-1}$.

The objective polynomial for the graph coloring problem is formulated as follows.
Let $V$ and $E$ be the sets of vertices and edges in the graph $G$.
Suppose that all the colors are $0,\cdots, 2^m-1$ for some $m\ge1$, and the color of the vertex $v\in V$ is represented by the binary representation $[x_{v,0}\cdots x_{v,m-1}]_{(2)}$.
Given two adjacent vertices $v$ and $w$, we define the polynomial $P_{v,w}$ as
\begin{equation}
	P_{v,w} = \prod_{k=0}^{m-1}(x_{v,k}x_{w,k} + (1-x_{v,k})(1-x_{w,k})).
\end{equation}
The value of $P$ is $0$ if and only if $x_{v,k} \neq x_{w,k}$ for some $k=0, \cdots, m-1$, i.e. when the colors of two vertices $v,w$ are different.
We define the polynomial $Q_G$ as
\begin{equation}\label{eqn:utility-polynomial}
	Q_G = \sum_{(v,w)\in E}P_{v,w}.
\end{equation}
The polynomial $Q_G$ is the objective function for the graph coloring problem on $G$.
The minimum value of $Q_G$ is $0$ and it is attained if and only if when the graph $G$ is properly colored, i.e. the coloring on $G$ satisfies the criteria of the vertex coloring.
With sufficiently large $m>0$, the solution to $Q_G=0$ always exists, while the real question is how small the $m$ could be in order to have such solution.

When $m$ is small, $Q_G=0$ may not be attainable.
Even though, finding the solution for the global minimum of $Q_G$ could be necessary, because it gives the optimal vertex coloring of the graph in the sense that it minimizes the number of conflicts, i.e. pairs of adjacent vertices that are colored with the same color.

The degree of the $Q_G$ is greater than two, except when there are only two colors in the color set.
Thus, in general, we need degree reduction methods to formulate the graph coloring problem as QUBO.
There are degree reduction methods that reduces monomials of degree greater than $2$ to a quadratic polynomial.
Freedman's method \cite{Fr05} applies to the monomials whose coefficients are negative:
\begin{align}\label{eqn-freed}
-\prod_{j=0}^{d-1} x_j &= \min_{w=0,1}w\left((d-1)-\sum_{j=0}^{d-1}x_j\right),
\end{align}
Ishikawa's method \cite{Is11} applies to the monomials whose coefficients are positive, and it has two distinct formulae depending on whether the degree of the monomial is even or odd:
\begingroup\allowdisplaybreaks
\begin{alignat}{2}
\prod_{j=0}^{2d+1} x_j 
	&=  \sum_{0\le i<j\le 2d+2} x_ix_j &&+\min_{\substack{w_i=0,1\\i=0,\ldots,d-1}} \left[\sum_{i=0}^{d-1}w_i\left((4i+3)- 2\sum_{j=0}^{2d+1}x_j\right) \right], \label{eqn-ishi-even}\\
\prod_{j=0}^{2d} x_j 
	&= \sum_{0\le i<j\leq2d} x_ix_j &&+\min_{\substack{w_i=0,1\\i=0,\ldots,d-1}} \left[\sum_{i=0}^{d-2}w_i\left((4i+3)- 2\sum_{j=0}^{2d}x_j\right) \right. \label{eqn-ishi-odd}\\
	& &&\qquad\qquad\qquad + \left.w_{d-1}\left((2d-1) - \sum_{j=0}^{2d}x_j\right)\right]. \nonumber
\end{alignat}
\endgroup

From now on, we will call these degree reduction methods as the \textbf{monomial reduction methods}. 
As an example, let us formulate the objective polynomial of the graph $G$ below, and apply the monomial reduction methods.\\
\begin{center}
\begin{tikzpicture}
\begin{scope}
	\node (G) at (-1,0) {$G:$};
\end{scope}
\begin{scope}[every node/.style={circle,thick,draw}]
	\node (0) at (1,0) {$0$};
	\node (1) at (3,0) {$1$};
	\node (2) at (5,0) {$2$};
\end{scope}
\begin{scope}[every node/.style={fill=white,circle}]
	\path (0) edge (1);
	\path (1) edge (2);
\end{scope}
\end{tikzpicture}
\end{center}
Let the colors at the vertices $0,1,2$ be denoted by $[x_1x_0]_{(2)}$, $[x_3x_2]_{(2)}$, $[x_5x_4]_{(2)}$, respectively.
Then objective function $Q_G$ is 
\begingroup
\allowdisplaybreaks
\begin{align}
Q_G &= (x_1x_3+(1-x_1)(1-x_3))(x_0x_2+(1-x_0)(1-x_2)) \label{eqn-ex-qg}\\
&\qquad + (x_3x_5+(1-x_3)(1-x_5))(x_2x_4+(1-x_2)(1-x_4)) \nonumber\\
&= 2 - x_0 - x_1 - 2x_2 - 2x_3 - x_4 - x_5 + x_0x_1 + 2x_0x_2 + x_0x_3 \nonumber\\
&\qquad + x_1x_2 + 2x_1x_3 + 2x_2x_3 + 2x_2x_4 + x_2x_5 + x_3x_4 + 2x_3x_5 \nonumber\\
&\qquad - 2x_0x_1x_2 - 2x_0x_1x_3 - 2x_0x_2x_3 - 2x_1x_2x_3 \nonumber\\
&\qquad - 2x_2x_3x_4 - 2x_2x_3x_5 - 2x_2x_4x_5 - 2x_3x_4x_5 \nonumber \\
&\qquad + 4x_0x_1x_2x_3 + 4x_2x_3x_4x_5 \nonumber
\end{align}
\endgroup

Let us apply Freedman's methods (c.f. Equations \eqref{eqn-freed}) to all monomials of the degree $3$:
\begingroup
\allowdisplaybreaks
\begin{align*}
- 2x_0x_1x_2 &= 2x_0x_1 + 2x_0x_2 + 2x_1x_2 + 2\min_{w_0=0,1}\left[3w_0 - x_0w_0 - x_1w_0 - x_2w_0\right]\\
- 2x_0x_1x_3 &= 2x_0x_1 + 2x_0x_3 + 2x_1x_3 + 2\min_{w_1=0,1}\left[3w_1 - x_0w_1 - x_1w_1 - x_3w_1\right]\\
- 2x_0x_2x_3 &= 2x_0x_2 + 2x_0x_3 + 2x_2x_3 + 2\min_{w_2=0,1}\left[3w_2 - x_0w_2 - x_2w_2 - x_3w_2\right]\\
- 2x_1x_2x_3 &= 2x_1x_2 + 2x_1x_3 + 2x_2x_3 + 2\min_{w_3=0,1}\left[3w_3 - x_1w_3 - x_2w_3 - x_3w_3\right]\\
- 2x_2x_3x_4 &= 2x_2x_3 + 2x_2x_4 + 2x_3x_4 + 2\min_{w_4=0,1}\left[3w_4 - x_2w_4 - x_3w_4 - x_4w_4\right]\\
- 2x_2x_3x_5 &= 2x_2x_3 + 2x_2x_5 + 2x_3x_5 + 2\min_{w_5=0,1}\left[3w_5 - x_2w_5 - x_3w_5 - x_5w_5\right]\\
- 2x_2x_4x_5 &= 2x_2x_4 + 2x_2x_5 + 2x_4x_5 + 2\min_{w_6=0,1}\left[3w_6 - x_2w_6 - x_4w_6 - x_5w_6\right]\\
- 2x_3x_4x_5 &= 2x_3x_4 + 2x_3x_5 + 2x_4x_5 + 2\min_{w_7=0,1}\left[3w_7 - x_3w_7 - x_4w_7 - x_5w_7\right]
\end{align*}
\endgroup

Let us apply Ishikawa's method (c.f Equation \eqref{eqn-ishi-even}) to all monomials of the degree $4$:
\begin{align*}
4x_0x_1x_2x_3 &= 4x_0x_1 + 4x_0x_2 + 4x_0x_3 + 4x_1x_2 + 4x_1x_3 + 4x_2x_3 \\
&\qquad + 4\min_{w_8=0,1}\left[3w_8 - 2x_0w_8 - 2x_1w_8 - 2x_2w_8 - 2x_3w_8\right]\\
4x_2x_3x_4x_5 &= 4x_2x_3 + 4x_2x_4 + 4x_2x_5 + 4x_3x_4 + 4x_3x_5 + 4x_4x_5 \\
&\qquad + 4\min_{w_9=0,1}\left[3w_9 - 2x_2w_9 - 2x_3w_9 - 2x_4w_9 - 2x_5w_9\right]
\end{align*}

Thus $Q_G$ is reduced to the following polynomial:
\begin{align*}
Q_G &= 2 - x_0 - x_ 1 - 2x_2 - 2x_3 - x_4 - x_5 + 4w_0 + 4w_1 + 4w_2 + 4w_3 + 4w_4 \\
&\qquad + 4w_5 + 4w_6 + 4w_7 + 12w_8 + 12w_9 + 5x_0x_1 + 6x_0x_2 + 5x_0x_3 + 5x_1x_2  \\
&\qquad + 6x_1x_3 + 10x_2x_3 + 6x_2x_4 + 5x_2x_5 + 5x_3x_4 + 6x_3x_5 + 5x_4x_5 \\
&\qquad +\min_{\substack{w_i=0,1\\i=0,\ldots,9}}\left[
\begin{array}{l}
 - 2x_0w_0 - 2x_1w_0 - 2x_2w_0 - 2x_0w_1 - 2x_1w_1 - 2x_3w_1 - 2x_0w_2 \\
- 2x_2w_2 - 2x_3w_2 - 2x_1w_3 - 2x_2w_3 - 2x_3w_3 - 2x_2w_4 - 2x_3w_4 \\
- 2x_4w_4 - 2x_2w_5 - 2x_3w_5 - 2x_5w_5 - 2x_2w_6 - 2x_4w_6 - 2x_5w_6 \\
- 2x_3w_7 - 2x_4w_7 - 2x_5w_7 - 8x_1w_8 - 8x_2w_8 - 8x_3w_8 - 8x_2w_9 \\
- 8x_3w_9 - 8x_4w_9 - 8x_5w_9
\end{array}\right]
\end{align*}

Note that $10$ auxiliary variables are produced, and they even outnumber the original variables.
This situation only gets worse when the number of vertices or edges increases. 
For example, the objective polynomial $Q_G$ for the complete graph $G=K_8$ with eight colors ($m=3$) consists of $24$ binary variables and has $1429$ distinct monomials. 
The monomial degree reduction on $Q_G$ produces $1156$ new auxiliary variables and there are total $5849$ distinct monomials in the reduced quadratic polynomial (c.f. Table \ref{tab:avg-num} in $\S\ref{sec:simulate}$).

Meanwhile, the objective function for the graph coloring problem may contain some homogeneous symmetric polynomials. 
For example, $Q_G$ in Equation \eqref{eqn-ex-qg} contains the following homogeneous symmetric polynomial 
\begin{equation}\label{eqn:p43-0-3}
-2P_4^{(3)}(x_0,x_1,x_2,x_3) = -2(x_0x_1x_2 + x_0x_1x_3 + x_0x_2x_3 + x_1x_2x_3). 
\end{equation}
Higher the connectivity of the graph $G$ is, larger the homogeneous symmetric polynomial that the objective function $Q_G$ contains.

The values of this symmetric polynomial only depends on the number of $1$'s in the binary variables.
For example, if $l=x_0+\cdots+x_3$, then $P_4^{(3)}(x_0,\cdots,x_3)=0$ for $l<3$, and $P_4^{(3)}(x_0,\cdots,x_3)=\displaystyle{l\choose3}$ for $3\le l\le 4$.
We can easily observe that this simple rule applies to all homogeneous symmetric polynomials of binary variables.
Let $S_n^{(m)}$ be the set of all subsets in $S_n=\{0,\cdots,n-1\}$ whose cardinality is $m$.
We will denote $P_n^{(m)}(x_0,\cdots,x_{n-1})$ as the homogeneous symmetric polynomial of degree $m$ on 
$n$ variables $x_0,\cdots,x_{n-1}$:
\begin{equation}\label{eqn-p-n-m-def}
P_n^{(m)}(x_0,\cdots,x_{n-1}) = \sum_{\{i_0,\cdots,i_{m-1}\}\in S_n^{(m)}}x_{i_0}\cdots x_{i_{m-1}}.
\end{equation}
Then, we have
\begin{equation}\label{defn-p_n_m}
P_n^{(m)}(x_0,\cdots,x_{n-1}) = 	\left\{\begin{array}{cl}
								0 & \textrm{if } \displaystyle\sum_{j=0}^{n-1}x_j< m \\
								\displaystyle{l\choose m} & \textrm{if } l = \displaystyle\sum_{j=0}^{n-1}x_j\ge m
								\end{array}\right.
\end{equation}

In this paper, we propose a new degree reduction method called the \textbf{symmetric reduction method} that reduces $P_n^{(m)}$ to a quadratic polynomial.
It produces less number of auxiliary variables than the monomial reduction on the objective polynomials for the graph coloring problem.
In $\S\ref{sec:main-results}$, we state main theorems \ref{thm-main-1}, \ref{thm-main-2} on the formulae of the reduced polynomials of the symmetric polynomials with positive and negative coefficients.
In $\S\ref{sec-proof-1},\ref{sec-proof-2}$, we prove these theorems.
In $\S\ref{sec:simulate}$, we describe algorithms \ref{tab:max-symm-alg}, \ref{tab:symm-red-alg} that implement the symmetric methods for applications .
In the same section, we present the results on testing these algorithms on random and complete graphs with various vertex sizes.
We compared the efficiency of the symmetric reduction method with the conventional reduction methods by comparing the numbers of variables and monomials of reduced polynomials obtained by two methods.

\section{The main results}\label{sec:main-results}

Let us re-define a binomial coefficient symbol $\displaystyle{n \choose m}$ as follows: for any two non-negative integers $n$ and $m$,
\begin{equation}\label{eqn-n-choose-m-def}
	{n \choose m}  = \left\{	\begin{array}{cl}
						\displaystyle\frac{n!}{m!(n-m)!} & \textrm{if } m \le n \\
						0 & \textrm{if } m>n
						\end{array}.\right.
\end{equation}
Here we state our two main theorems.
\begin{thm}\label{thm-main-1}
Let $n\ge m>2$ and define $L = \displaystyle {n-2\choose m-2},\quad d = \displaystyle\left\lfloor\frac{n-1}{2}\right\rfloor$.
For each $0\le i\le d-1$, let us define
\begin{align}
a_i &= (4i+3){n-2\choose m-2} - (m-1)\left({2i+3\choose m} - {2i+1\choose m} \right),\label{eqn-thm-main-1-a-i-2d-1}\\
b_i &= {2i+2\choose m-1} - {2i\choose m-1} - 2{n-2\choose m-2}.\label{eqn-thm-main-1-b-i-2d-1}
\end{align}
Then the following equality holds:
\begin{align}\label{eqn-p-n-m-pos-red}
P_n^{(m)}(\mathbf x) 
&= L\sum_{0\le i<j\le n-1}x_ix_j + \min_{\substack{w_i=0,1\\i=0,\cdots,d-1}}\left[\sum_{i=0}^{d-1} w_i\left(a_i + b_i\sum_{j=0}^{n-1} x_j\right)\right].
\end{align}
\end{thm}

\begin{thm}\label{thm-main-2}
Let $n\ge m>2$ and define $d = \displaystyle\left\lfloor\frac{n-m+2}{2}\right\rfloor$.
If $n-m = 2d-1$, then for each $0\le i\le d-1$,
\begin{align}
a_i &= (m-1)\left({m+2i+1\choose m} - {m+2i-1\choose m}\right),\label{eqn-thm-2-ai-odd}\\
b_i &= -{m+2i\choose m-1} + {m+2i-2\choose m-1} \textrm{ for }0\le i\le d-1 \label{eqn-thm-2-bi-odd}.
\end{align}
If $n-m = 2d-2$, then Equations \eqref{eqn-thm-2-ai-odd}, \eqref{eqn-thm-2-bi-odd} hold for $0\le i\le d-2$ and let
\begin{equation} 
	a_{d-1} = (m-1){n-1\choose m-1},\quad b_{d-1} = -{n-2\choose m-2}.
\end{equation}
Then the following equality holds:
\begin{equation}
-P_n^{(m)}(\mathbf x) = \min_{\substack{w_i=0,1\\i=0,\cdots,d-1}}\sum_{i=0}^{d-1} w_i\left(a_i + b_i\sum_{j=0}^{n-1}x_j\right).
\end{equation}
\end{thm}

Before we present the proofs for Theorems \ref{thm-main-1}, \ref{thm-main-2}, let us explain the underlying ideas with examples.

\begin{exmp}\label{exmp-p-5-3-pos}
First, let us explain Theorem \ref{thm-main-1} with $P_5^{(3)}(\mathbf x)$.
Our goal is to find a quadratic polynomial $Q$ that attains the same value of $P_5^{(3)}(\mathbf x)$ consists of variables $x_0,\cdots,x_4$ together with $d$ auxiliary variables $w_0,\ldots,w_{d-1}$. 
The only possible formula for $Q$ is 
\begin{equation}\label{eqn:qubo-form-p53}
    Q = \sum_{0\le i<j\le4} L_{ij}x_ix_j + \sum_{j=0}^4 c_jx_j
    	+ \min_{\substack{w_i=0,1\\i=0,\ldots,d-1}}\left[\sum_{i=0}^{d-1}w_i\left(a_i + \sum_{j=0}^4b_{i,j}x_j\right)\right].
\end{equation}
In fact, this could be the general form of any reduced polynomial if it had a constant term.
We omitted the constant term here, since we know that the minimum $P_5^{(3)}(\mathbf x)$ is $0$ (and the same for $P_n^{(m)}(\mathbf x)$ in general).
Just for now, let us assume that $b_{i,j}=b_i$, $L_{i,j} = L$, and $c_i=0$.
This means that $b_{i,j}$ will depend on the index of the auxiliary variables, $L_{i,j}$ is a constant, and $Q$ does not have any linear terms on $x_j$'s.

\begin{table}[h!]
\centering
\begin{tabular}{c|c|c|c|c|c}
$l(\mathbf x)$ & $P_5^{(3)}(\mathbf x) $ & $\displaystyle\sum_{i<j}x_ix_j$ & 
$A_l$ & $\displaystyle\min_{w_0=0,1}w_0(7-5l(\mathbf x))$ & 
$\displaystyle\min_{w_1=0,1}w_1(3-l(\mathbf x))$ \\
\hline\hline 
$0$ & $0$   & $0$   & $0$   & $0$   & $0$ \\
$1$ & $0$   & $0$   & $0$   & $0$   & $0$ \\
$2$ & $0$   & $1$   & $-3$  & $-3$  & $0$ \\
$3$ & $1$   & $3$   & $-8$  & $-8$  & $0$ \\
$4$ & $4$   & $6$   & $-14$ & $-13$ & $-1$ \\
$5$ & $10$  & $10$  & $-20$ & $-18$ & $-2$ \\
\end{tabular}
\caption{The sequence $A_l$ (c.f. Equation \eqref{eqn:al}) when $L=3$ and the arithmetic progressions that sums up to $A_l$.}
\label{tab:p53-al}
\end{table}

Let $l = l(\mathbf x)=x_0 + \ldots x_4$ and define the sequence $A_l$ by
\begin{equation}\label{eqn:al}
	A_l = P_5^{(3)}(\mathbf x) - L\sum_{0\le i<j\le 4}x_ix_j.
\end{equation}
The second and third columns of Table \ref{tab:p53-al} show the values for $P_5^{(3)}(\mathbf x)$ and $\displaystyle \sum_{0\le i,j\le 4}x_ix_j$ for $l = 0,\cdots, 5$.
If we choose $L=3$, then the sequence $A_l$ becomes a decreasing sequence of non-positive integers (the fourth column of Table \ref{tab:p53-al}).
In fact, $L=3$ is the minimum integer that makes the sequence $A_l$ as a decreasing sequence.
The first two non-zero numbers in $A_l$ are $-3, -8$ at $l=2, 3$ respectively, and they follows the arithmetic progression $7-5l$.
Let us subtract $\displaystyle \min_{w_0=0,1} w_0(7-5l)$ (the fifth column of Table \ref{tab:p53-al}) from $A_l$ for each $l$.
Then we get a sequence that has only two negatives $-1,-2$ at $l=4, 5$ respectively, where all other values are $0$ (the sixth column of Table \ref{tab:p53-al}). 
This sequence is the same as $\displaystyle\min_{w_1=0,1} w_1(3-l)$.
Thus we obtain the following identity:
\begin{equation}
	P_5^{(3)}(\mathbf x) = 3\sum_{0\le i<j\le 4}x_ix_j 
					+ \min_{w_0,w_1=0,1}\left[w_0(7-5\sum_{j=0}^4x_j) + w_1(3-\sum_{j=0}^4x_j)\right].
\end{equation}
It is worth noticing that we reduced the polynomial $P_5^{(3)}(\mathbf x)$ with only two auxiliary variables $w_0$ and $w_1$, whereas the monomial reductions would produce $10$ auxiliary variables, one for each monomial (c.f. Equation \eqref{eqn-ishi-odd}).
\end{exmp}

\begin{exmp}\label{exmp-p-3-5-neg}

\begin{table}[h!]
\centering
\begin{tabular}{c|c|c|c}
$l(\mathbf x)$ & $B_l$ &
$\displaystyle\min_{w_0=0,1}w_0(8-3l(\mathbf x))$ &
$\displaystyle\min_{w_1=0,1}w_1(12-3l(\mathbf x))$\\
\hline\hline 
$0$ & $0$ & $0$ & $0$ \\
$1$ & $0$ & $0$ & $0$ \\
$2$ & $0$  & $0$ & $0$ \\
$3$ & $-1$ & $-1$ & $0$ \\
$4$ & $-4$ & $-4$ & $0$ \\
$5$ & $-10$ & $-7$ & $-3$\\
\end{tabular}
\caption{The values of $B_l$ in Equation \eqref{eqn-p53-bl} and the arithmetic progressions that sums up to $B_l$.}
\label{tab:bl}
\end{table}

Next, let us explain Theorem \ref{thm-main-2} with the example of $-P_5^{(3)}(\mathbf x)$.
Let us define a sequence $B_l$ as 
	\begin{equation}\label{eqn-p53-bl}
	B_l = -P_5^{(3)}(\mathbf x).
	\end{equation}
The sequence $B_l$ is already a decreasing sequence of non-positive numbers, and its numbers are shown in the second column of Table \ref{tab:bl}.
Let us observe the first two non-positives $-1, -4$ at $l=3, 4$ respectively.
The arithmetic progression that matches with these numbers is $8-3l$.
We subtract the sequence $\displaystyle \min_{w_0=0,1} w_0(8-3l)$ from $B_l$ to obtain a sequence of zeros except $-3$ at $l=5$. 
Then we take the last two non-positive integers $0,-3$ at $l=4,5$ respectively, and use the arithmetic progression $12-3l$ to match these numbers.
In this way, we get the following reduction:
\begin{equation}
-P_5^{(3)}(\mathbf x) =\min_{w_0,w_1=0,1}\left[w_0(8-3\sum_{j=0}^4x_j) + w_1(12-3\sum_{j=0}^4x_j).\right]
\end{equation}
Again, we reduced the polynomial $-P_5^{(3)}(\mathbf x)$ with only two auxiliary variables whereas the monomial reductions would produce $10$ auxiliary variables (c.f. Equation \eqref{eqn-ishi-odd})
\end{exmp}

\begin{rem}\label{rem-imp}
Before we proceed to the next sections, let us point out some important facts.
\begin{enumerate}
\item In the sequence $A_l = P_n^{(m)}(\mathbf x) - L\cdot l(\mathbf x)$, the constant $L$ is set as minimum as possible to make $A_l$ a non-positive decreasing sequence.
There are $n-1$ negatives in $A_l$, so we need $\displaystyle\lfloor (n-1)/2\rfloor$ auxiliary variables in the reduced polynomial of $P_n^{(m)}(\mathbf x)$.
\item In the sequence $B_l = -P_n^{(m)}(\mathbf x)$, there are $n-m+2$ negatives.
Thus, we need $\displaystyle\lfloor(n-m+2)/2\rfloor$ auxiliary variables in the reduced polynomials of $-P_n^{(m)}(\mathbf x)$.
\item Ishikawa's formulae in Equations \eqref{eqn-ishi-even}, \eqref{eqn-ishi-odd} are the special cases of Theorem \ref{thm-main-1} when $n=m$.
If $n=2d+2$, then for all $0\le i\le d-1$, we have $2i+3 \le 2d+1< n=m$ and $2i+2\le 2d < m-1$, so Equations \eqref{eqn-thm-main-1-a-i-2d-1}, \eqref{eqn-thm-main-1-b-i-2d-1} become
\begin{equation}\label{eqn-a-i-b-i-special}
a_i = (4i+3),\quad b_i = -2.
\end{equation}
If $n=2d+1$, then Equation \eqref{eqn-a-i-b-i-special} hold for all $0\le i\le d-2$, and
$$a_{d-1} = (4d-1)-(n-1) = 2d-1\quad b_{d-1} = 1-2 = -1,$$
which coincides with Ishikawa's formula in Equations \eqref{eqn-ishi-even}, \eqref{eqn-ishi-odd}.
Freedman's formula in Equation \eqref{eqn-freed} is a direct consequence of Theorem \ref{thm-main-2} when $n=m$: there is only one pair of coefficients $a_0, b_0$, which are
$$a_0 = n-1,\quad b_0 = -1.$$
Thus, we can say that Theorems \ref{thm-main-1}, \ref{thm-main-2} are the generalizations of Ishikawa's and Freedman's methods.
\end{enumerate}
\end{rem}

\section{Proof of Theorem \ref{thm-main-1}}\label{sec-proof-1}

The original binomial coefficients satisfies the following property: for $0\le s< r$,
\begin{equation}\label{eqn:bin-prop}
	{r\choose s} + {r\choose s+1} = {r+1\choose s+1}.
\end{equation}
With the new definition of binomial coefficients in Equation \eqref{eqn-n-choose-m-def}, the same property holds for all $r,s\ge0$:
if $s= r$, both sides of Equation \eqref{eqn:bin-prop} are $1$; if $s\ge r+1$, both sides of Equation \eqref{eqn:bin-prop} are $0$.
We will use Equation \eqref{eqn:bin-prop} often in the subsequent calculations.

Let us fix the variables as $x_0,\cdots,x_{n-1}$ and omit the notation $\mathbf x = (x_0,\cdots,x_{n-1})$ whenever there is no ambiguity. 
Let us define the sequence $A_l$ as follows:
\begin{equation}\label{eqn-al-defn}
	A_l = P_n^{(m)} - L\displaystyle\sum_{0\le i<j\le n-1}x_ix_j.
\end{equation}

Our first task is to find a value $L$ so that the sequence $A_l$ satisfies the following two properties:
\begin{itemize}
	\item[\textbf{P1}] $A_l$ is a decreasing finite sequence of non-positive integers for $0\le l\le n$;
	\item[\textbf{P2}] the difference between two consecutive numbers in $A_l$ is also decreasing, i.e.
	\begin{equation}\label{eqn-a-l-diff-dec}
	(A_{n-i}-A_{n-i-1}) \le (A_{n-i-1}-A_{n-i-2})\textrm{ for all }0\le i\le n-2.
	\end{equation}
\end{itemize} 
For each $0\le i \le n-2$, Equation \eqref{eqn-al-defn} gives a triple of consecutive numbers in $A_l$:
\begin{align}
    A_{n-i-2} &= {n-i-2\choose m} - L{n-i-2\choose 2}, \label{eqnani1}\\
    A_{n-i-1} &= {n-i-1\choose m} - L{n-i-1\choose 2}, \label{eqnani2}\\
    A_{n-i} &= {n-i\choose m} - L{n-i\choose 2}.\label{eqnani3}
\end{align}
We can re-write the inequality \eqref{eqn-a-l-diff-dec} as
\begin{align*}
&\left({n-i\choose m}-{n-i-1\choose m}\right) - \left({n-i-1\choose m}-{n-i-2\choose m}\right) \\
&\qquad - L\left(\left({n-i\choose 2}-L{n-i-1\choose 2}\right) - \left({n-i-1\choose 2}-{n-i-2\choose 2}\right)\right)\le 0.
\end{align*}
Thus we have
\begin{align*}
	&\left({n-i-1\choose m-1} - {n-i-2\choose m-1}\right) - L\left({n-i-1\choose 1} - {n-i-2\choose 1}\right) \\
	&= {n-i-2\choose m-2} - L\le 0.
\end{align*}
The number of negatives in the sequence $A_l$ determines the numbers of auxiliary variables in the reduced polynomial (c.f. Examples \ref{exmp-p-5-3-pos}, \ref{exmp-p-3-5-neg} and Remark \ref{rem-imp}).
Thus $L$ should be chosen as minimum as possible so that $A_l$ has the least number of negatives.
Therefore,
\begin{equation}\label{eqn:llnm}
	L = \displaystyle{n-2\choose m-2}.
\end{equation}
With this $L$, all $A_l$'s are also are non-positive (i.e. $A_l$ satisfies \textbf{P1}).

Next, we want to find the coefficients $a_j, b_j$ that satisfy
\begin{equation}\label{eqn:al-w-1-a-b}
	A_l = P_n^{(m)} - L{l\choose 2} = \min_{\substack{w_i=0,1\\i=0,\cdots,d-1}}\left[\sum_{i=0}^{d-1} w_i\left(a_i + b_i l\right)\right].
\end{equation}
There are $n-1$ negative integers in $A_l$ at $l=2,\cdots, n$.
We need to split the cases for $n$ where $n$ is odd or $n$ is even.

%
\subsection{The odd case}
Suppose that $n=2d+1$ for some $d$. 
Then there are $2d$ negatives in $A_l$ and we want $d$ arithmetic progressions in Equation \eqref{eqn:al-w-1-a-b} subsequently appear as follows:
\begingroup
\allowdisplaybreaks
\begin{align}
	& {2\choose m} - {n-2\choose m-2}{2 \choose 2} = a_0+2b_0, \label{eqn-thm-1-odd-2-m}\\
	& {3\choose m} - {n-2\choose m-2}{3 \choose 2} = a_0+3b_0, \quad (w_0=1, w_k=0\textrm{ for }1\le k\le d-1) \label{eqn-thm-1-odd-3-m}\\
	& {4\choose m} - {n-2\choose m-2}{4 \choose 2} = (a_0+4b_0) + (a_1+4b_1), \label{eqn-thm-1-odd-4-m}\\
	& {5\choose m} - {n-2\choose m-2}{5 \choose 2} = (a_0+5b_0) + (a_1+5b_1),\label{eqn-thm-1-odd-5-m}\\
	& \qquad  \qquad \qquad \qquad \qquad \qquad\qquad (w_0=w_1=1, w_k=0\textrm{ for }2\le k\le d-1) \nonumber \\
	&\qquad\vdots\nonumber\\
	& {n-1\choose m} - {n-2\choose m-2}{n-1\choose 2} = \sum_{k=0}^{d-1} a_k + (n-1)b_k,\label{eqn-thm-1-odd-n-1-m}\\
	& {n\choose m} - {n-2\choose m-2}{n\choose 2} = \sum_{k=0}^{d-1}a_k + nb_k.\quad (w_0=\cdots=w_{d-1}=1)\label{eqn-thm-1-odd-n-m}
\end{align}
\endgroup
That is, for each $0\le i \le d-1$,
\begin{align}
	& {2i+2\choose m} - {n-2\choose m-2}{2i+2\choose 2} = \sum_{k=0}^{i} a_k + (2i+2)b_k,\label{eqn-thm-1-odd-2i+2-m}\\
	& {2i+3\choose m} - {n-2\choose m-2}{2i+3\choose 2} = \sum_{k=0}^{i} a_k + (2i+3)b_k.\label{eqn-thm-1-odd-2i+3-m}
\end{align}
This is possible because the sequence $A_l$ satisfies the inequality \eqref{eqn-a-l-diff-dec}: the differences $A_{l+1}-A_l$ are getting smaller (as negatives) as $l$ increases.
We equate $A_l$ on the left hand-side of Equation \eqref{eqn:al-w-1-a-b} by introducing new arithmetic progressions of negative common differences on the right hand-side.

To find the formulae of $b_i$, we subtract Equations \eqref{eqn-thm-1-odd-2i+2-m} from Equation \eqref{eqn-thm-1-odd-2i+3-m} to get
\begingroup
\allowdisplaybreaks
\begin{align}
	\sum_{k=0}^{i}b_k 
	&= {2i+3 \choose m} - {2i+2 \choose m} - {n-2 \choose m-2} \left({2i+3\choose 2}- {2i+2\choose 2}\right) \nonumber\\
	&= {2i+2 \choose m-1} - {n-2 \choose m-2}((2i+3)(i+1) - (i+1)(2i+1)) \nonumber\\
	&= {2i+2 \choose m-1} - 2(i+1){n-2 \choose m-2} \label{eqn-thm-1-sum-b-k-odd}
\end{align}
\endgroup
Thus for $0\le i\le d-1$,
\begin{align}
	b_i 	&= \left({2i+2 \choose m-1} - 2(i+1){n-2 \choose m-2}\right) - \left({2i \choose m-1} - 2i{n-2 \choose m-2}\right) \nonumber\\
		&= {2i+2\choose m-1} - {2i\choose m-1} - 2{n-2\choose m-2}.\label{eqn-thm-1-b-i-odd}
\end{align}

To find the formula of $a_i$, we add Equations \eqref{eqn-thm-1-odd-2i+2-m}, \eqref{eqn-thm-1-odd-2i+3-m} to get
\begin{align*}
	2\sum_{k=0}^{i}a_k + (4i+5)\sum_{k=0}^{i}b_k 
	&={2i+2\choose m} + {2i+3\choose m} \\
	&\qquad - ((2i+3)(i+1)+(i+1)(2i+1)){n-2\choose m-2}  
\end{align*}
Applying Equation \eqref{eqn-thm-1-sum-b-k-odd}, we get
\begingroup
\allowdisplaybreaks
\begin{align}
	\sum_{k=0}^{i}a_k 
	&= \frac{1}{2}\left({2i+3\choose m} + {2i+2\choose m} - 4(i+1)^2{n-2\choose m-2}\right. \label{eqn-thm-1-sum-a-k}\\
	&\qquad \left.- (4i+5)\left({2i+2 \choose m-1} - 2(i+1){n-2 \choose m-2}\right)\right) \nonumber \\
	&= \frac{1}{2}\left({2i+3\choose m} + {2i+2\choose m} - (4i+5){2i +2 \choose m-1}\right. \nonumber \\
	&\qquad \left.+ ((4i+5)(2(i+1))-4(i+1)^2){n-2\choose m-2}\right) \nonumber \\
	&= \frac{1}{2}\left({2i+3\choose m} + \left({2i+2\choose m} + {2i + 2 \choose m-1}\right) - 2(2i+3){2i +2 \choose m-1}\right. \nonumber\\
	&\qquad \left. + 2(2i^2+5i+3){n-2\choose m-2}\right) \nonumber\\
	&= (i+1)(2i+3){n-2\choose m-2} - (m-1){2i+3\choose m}, \nonumber\\
	a_i &= \left((i+1)(2i+3)-i(2i+1)\right){n-2\choose m-2} \label{eqn-thm-1-a-i-odd}\\
	&\qquad - (m-1)\left({2i+3\choose m} - {2i+1\choose m} \right) \nonumber \\
	&= (4i+3){n-2\choose m-2} - (m-1)\left({2i+3\choose m} - {2i+1\choose m} \right)\nonumber
\end{align}
\endgroup

%
\subsection{The even case}
Suppose that $n = 2d+2$.
We have $2d-1$ negatives in the sequence $A_l$ for starting from $l = 2$.
We can write the first few equations of Equation \eqref{eqn:al-w-1-a-b} exactly as Equations \eqref{eqn-thm-1-odd-2-m} -- \eqref{eqn-thm-1-odd-5-m}.
However, the last three equations will be
\begin{align}
	{n-2\choose m} - {n-2\choose m-2}{n-2\choose 2} &= \sum_{k=0}^{d-1} a_k + (n-2)b_k,\label{eqn-thm-1-even-n-2-m}\\
	{n-1\choose m} - {n-2\choose m-2}{n-1\choose 2} &= \sum_{k=0}^{d-1} a_k + (n-1)b_k,\label{eqn-thm-1-even-n-1-m}\\
	{n\choose m} - {n-2\choose m-2}{n\choose 2} &= \sum_{k=0}^{d-1} a_k + nb_k\quad (w_0=\cdots=w_{d-1}=1).\label{eqn-thm-1-even-n-m}	
\end{align}
This happens because the triple $A_{n-2}, A_{n-1},A_n$ are always an arithmetic progression with a common difference:
\begingroup
\allowdisplaybreaks
\begin{align*}
A_{n-1}-A_{n-2} 
	&= \left({n-2\choose m} - {n-2\choose m-2}{n-1\choose 2}\right) - \left({n-2\choose m} - {n-2\choose m-2}{n-2\choose 2}\right)\\
	&= \left({n-1\choose m} - {n-2\choose m}\right) - {n-2\choose m-2}\left({n-1\choose 2} - {n-2\choose 2}\right) \\
	&= {n-2\choose m-1} - (n-2){n-2\choose m-2} \\
	&= {n-2\choose m-1} + {n-2\choose m-2} - (n-1){n-2\choose m-2} \\
	&= {n-1\choose m-1} - (m-1){n-1\choose m-1} \\
	&= -(m-2){n-1\choose m-1},\\
A_n - A_{n-1}
	&= \left({n\choose m} - {n-2\choose m-2}{n\choose 2}\right) - \left({n-1\choose m} - {n-2\choose m-2}{n-1\choose 2}\right) \\
	&= \left({n\choose m} - {n-1\choose m}\right) - {n-2\choose m-2}\left({n\choose 2} - {n-1\choose 2}\right) \\
	&= {n-1\choose m-1} - (n-1){n-2\choose m-2} \\
	&= {n-1\choose m-1} - (m-1){n-1\choose m-1} \\
	&= -(m-2){n-1\choose m-1}
\end{align*}
\endgroup
Therefore, the same formulae of $a_i$ and $b_i$ for $0\le i\le d-1$ in Equations \eqref{eqn-thm-1-b-i-odd}, \eqref{eqn-thm-1-a-i-odd} hold.
\section{Proof of Theorem \ref{thm-main-2}}\label{sec-proof-2}
Let us define the sequence $B_l$ as follows:
\begin{equation}\label{eqn-thm-2-b-l}
	B_l = -P_n^{(m)}
\end{equation}
The sequence $B_l$ satisfies the properties \textbf{P1}, \textbf{P2} in the previous section: first, each $B_l$ is obviously non-positive. 
Second, if $n\ge m+2$, then for each $0\le i\le n-m-2$, the triple of consecutive numbers $B_{m+i}, B_{m+i+1}, B_{m+i+2}$ satisfies
\begin{align*}
	&(B_{m+i+2} - B_{m+i+1}) - (B_{m+i+1} - B_{m+i}) \\
	&=-{m+i+2\choose m} + {m+i+1\choose m} - \left(-{m+i+1\choose m} + {m+i\choose m}\right) \\
	&=-{m+i+1\choose m-1} + {m+i\choose m-1} = -{m+i\choose m-2} < 0 
\end{align*}

The goal is to find the coefficients $a_i, b_i$ satisfying 
\begin{equation}\label{eqn-thm-2-b-l-min}
	B_l = -{l\choose m} = \min_{\substack{w_i = 0,1\\i=0,\cdots,d-1}}\left[\sum_{i=0}^{d-1}w_i\left(a_i+b_il\right)\right].
\end{equation}
There are $n+1-m$ negative numbers in the sequence $B_l$ for $m\le l\le n$.
Thus we need to split the cases for $n$ when $n-m$ is odd or even.

\subsection{The odd case} 
Suppose that $n-m = 2d-1$, i.e. $n-m+1 = 2d$.
Then there are $d$ arithmetic progressions in Equation \eqref{eqn-thm-2-b-l-min} that appears subsequently as follows: 
\begingroup
\allowdisplaybreaks
\begin{align}
	- {m\choose m} &= a_0+mb_0, \label{eqn-thm-2-odd-m-m}\\
	- {m+1\choose m} &=a_0+(m+1)b_0,\quad (w_0=1,w_k=0\textrm{ for }1\le k\le d-1)\label{eqn-thm-2-odd-m+1-m}\\
	- {m+2\choose m} &= (a_0+(m+2)b_0) + (a_1+(m+2)b_1), \label{eqn-thm-2-odd-m+2-m}\\
	- {m+3\choose m} &= (a_0+(m+3)b_0) + (a_1+(m+3)b_1, \label{eqn-thm-2-odd-m+3-m}\\
	&\qquad\qquad\qquad(w_0=w_1=1,w_k=0\textrm{ for }2\le k\le d-1)\nonumber\\
	&\vdots \nonumber\\
	-{n-1\choose m} &= \sum_{k=0}^{d-1}a_k + (n-1)b_k, \label{eqn-thm-2-odd-n-1-m}\\
	-{n\choose m} &= \sum_{k=0}^{d-1}a_k + nb_k.\quad (w_0=\cdots=w_{d-1}=1) \label{eqn-thm-2-odd-n-m}
\end{align}
\endgroup
For $0\le i\le d-1$, we have
\begin{align}
	-{m+2i\choose m} &= \sum_{k=0}^{i}a_k + (m+2i)b_k, \label{eqn-thm-2-odd-m+2i-m}\\
	-{m+2i+1\choose m} &= \sum_{k=0}^{i}a_k + (m+2i+1)b_k, \label{eqn-thm-2-odd-m+2i+1-m}
\end{align}
Subtracting Equation \eqref{eqn-thm-2-odd-m+2i-m} from Equation \eqref{eqn-thm-2-odd-m+2i+1-m}, we get 
\begin{equation}\label{eqn-thm-2-odd-sum-b-k}
	\sum_{k=0}^i b_k = -{m+2i+1\choose m} + {m+2i\choose m} = -{m+2i\choose m-1}.
\end{equation}
Therefore, for $0\le i\le d-1$,
\begin{equation}
	b_i = -{m+2i\choose m-1} + {m+2i-2\choose m-1}.
\end{equation}

By adding Equations \eqref{eqn-thm-2-odd-m+2i-m}, \eqref{eqn-thm-2-odd-m+2i+1-m}, we get 
\begin{equation}
2\sum_{k=1}^ia_k + (2m+4i+1)\sum_{k=1}^ib_k = -{m+2i+1\choose m} - {m+2i\choose m}. 
\end{equation}
Applying Equation \eqref{eqn-thm-2-odd-sum-b-k}, we get
\begin{align*}
    \sum_{k=0}^ia_k 
    &= \frac{1}{2}\left(-{m+2i+1\choose m} - {m+2i\choose m} + (2m+4i+1){m+2i\choose m-1}\right) \\
    &= \frac{1}{2}\left(- \left({m+2i+1\choose m} + {m+2i\choose m} + {m+2i\choose m-1}\right) + 2(m+2i+1){m+2i\choose m-1}\right) \\
    &= -{m+2i+1\choose m} + (m+2i+1){m+2i\choose m-1} \\
    &= (m-1){m+2i+1\choose m}
\end{align*}
Therefore, 
\begin{equation}
a_i = (m-1)\left({m+2i+1\choose m} - {m+2i-1\choose m}\right)
\end{equation}

%
\subsection{The even case}
Next, suppose that $n-m = 2d-2$.
In this case, the first few equations of Equation \eqref{eqn-thm-2-b-l-min} are exactly same as Equations \eqref{eqn-thm-2-odd-m-m} -- \eqref{eqn-thm-2-odd-m+3-m}, but the last three equations would be
\begin{align}
	-{n-2\choose m} &= \sum_{k=0}^{d-2}a_k + (n-2)b_k, \label{eqn-thm-2-even-n-2-m}\\
	-{n-1\choose m} &= \sum_{k=0}^{d-2}a_k + (n-1)b_k, \quad (w_0=\cdots=w_{d-2}=1)\label{eqn-thm-2-even-n-1-m}\\
	-{n\choose m} &= \sum_{k=0}^{d-2}\left(a_k + nb_k\right) + (a_{d-1} + nb_{d-1}). \quad (w_{d-1}=1)\label{eqn-thm-2-even-n-m}
\end{align}
The last arithmetic progression $a_{d-1} + b_{d-1}l$ must satisfy $a+(n-1)b_{d-1} = 0$.
This means 
\begin{align}
b_{d-1} &= - {n\choose m} + {n-1\choose m} - \sum_{k=0}^{d-2}b_k \label{eqn-b-d-1-even-4}\\
&= - {n-1\choose m-1} + {m + 2(d-2)\choose m-1} = - {n-1\choose m-1} + {n-2\choose m-1} \nonumber\\
&= - {n-2\choose m-2},\nonumber\\
a_{d-1} &= -(n-1)b_d = (n-1){n-2\choose m-2} \label{eqn-a-d-1-evne-4}\\
&= (m-1){n-1\choose m-1}.\nonumber
\end{align}

\section{Simulated results}\label{sec:simulate}

In this section, we introduce two algorithms, \textbf{MaxSymm} and \textbf{SymmRed}, which implement the symmetric degree reduction method based on Theorems \ref{thm-main-1}, \ref{thm-main-2}.
\textbf{MaxSymm} is an algorithm that finds \emph{maximal} (which will be defined shortly) symmetric polynomials in a given polynomial.
\textbf{SymmRed} is an algorithm that obtains a reduced quadratic polynomial from a given polynomial of binary variables, which prioritize the degree reduction of symmetric polynomials found by \textbf{MaxSymm}.
After reducing all symmetric homogeneous polynomials, \textbf{SymmRed} still uses the monomial reduction methods for reducing the remaining monomials.

We will call $P_n^{(m)}(\mathbf x)$ for $\mathbf x=(x_0,\cdots,x_{n-1})$ in Equation \eqref{defn-p_n_m} as the \textbf{symmetric $m$-polynomial on $\mathbf x$}.
We will say a polynomial $q$ \textbf{lies in} a polynomial $p$ if all monomials in $q$ are monomials in $p$.
In this case, we will say $p$ \textbf{contains} $q$.
A symmetric $m$-polynomial $q$ is said to be \textbf{maximal in $p$} if $q$ lies in $p$ and there is no other symmetric $m$-polynomial that lies in $p$ and contains $q$.
For example, the symmetric $3$-polynomial on $5$ variables $P_5^{(3)}(x_0,\cdots,x_4)$ is larger than the symmetric $3$-polynomial on $4$ variables $P_4^{(3)}(x_0,\cdots,x_3)$.

The \textbf{MaxSymm} algorithm takes two arguments, the polynomial $p$ and the degree $m$.
Again, this algorithm finds maximal symmetric $m$-polynomials that lies in $p$.
We say a symmetric $m$-polynomial is \textbf{trivial} if it consists of only $m$ variables, and the algorithm returns nil if there is no symmetric $m$-polynomial except trivial ones.
Let us call a monomial of degree $m$ simply as \textbf{$m$-monomial}.
We can represent every $m$-monomial by a set of distinct $m$ variables (or more simply, $m$ indices).
Generally, we can represent every symmetric $m$-polynomial by a set of distinct variables (or indices) whenever there is no ambiguity on the constant $m$.
For example, the symmetric polynomial $P_5^{(3)}(x_0,\cdots,x_4)$ is characterized by the set $\{x_0,x_1,x_2,x_3,x_4\}$, or more simply, $\{0,1,2,3,4\}$.

\begin{algorithm}
\caption{\textbf{MaxSymm} algorithm}
\label{tab:max-symm-alg}
\begin{algorithmic}[1]
\Procedure{MaxSymm}{$p$, $m$}\Comment{$p$ is a polynomial, $m\ge3$ is a degree}
\State $A,B\gets$ the set of all $m$-monomials in $p$
\While{$C$ is not empty}
	\State $B,C\gets C,\emptyset$ \label{alg-b-c-symm}
	\For{$q$ in $B$} \label{alg-symm-rb}
		\For{$r$ in $A$} \label{alg-symm-ra}
			\If{$r\not\subset q$}\label{alg-r-not-in-q}
				\State $s\gets$ the symmetric $m$-polynomial on $q\cup r$ \label{alg-symm-mp}
				\If{$s$ lie in $p$}\label{alg-s-in-p}
					\State{$C\gets C\cup\{s\}$} \label{alg-b-c-c-a}
				\EndIf
			\EndIf
		\EndFor
	\EndFor
\EndWhile
\State \textbf{return} $B$
\EndProcedure
\end{algorithmic}
\end{algorithm}

Algorithm \ref{tab:max-symm-alg} is a pseudocode ot \textbf{MaxSymm} algorithm.
It utilizes three sets $A, B, C$.
The set $A$ is the set of the all $m$-monomials in $p$, and it is the reference set so we do not change it throughout the algorithm.
The set $B$ is the set of largest symmetric $m$-polynomials in $p$ polynomials that are currently found.
The set $C$ is the set of all symmetric $m$-polynomials in $p$ that are even larger than at least one in $B$.
The algorithm starts by initializing the sets $A, B, C$ as the set of all monomials in $p$.
The initialization of $C$ is only for the first iteration of the below cycle to happen.
(The line numbers refers to Algorithm \ref{tab:max-symm-alg}.)
\begin{enumerate}
	\item Firstly, we replace the set $B$ by $C$ and empty the set $C$ (Line \ref{alg-b-c-symm}).
	This means we transfer all information on the symmetric $m$-polynomials in $C$ to $B$.
	\item Secondly, we look up all pairs $(q,r)$ of monomials in $B$ and $A$ (Lines \ref{alg-symm-rb}, \ref{alg-symm-ra}). 
	Only when $r$ contains an element which is not in $q$ (Line \ref{alg-r-not-in-q}), the algorithm creates the symmetric $m$-polynomial $s$ on all variables in the set $q\cup r$ (Line \ref{alg-symm-mp}).
	If $s$ lies in $p$, we insert $s$ in $C$ as an element (Lines \ref{alg-s-in-p}, \ref{alg-b-c-c-a}).
	\item The cycle aborts when there is no update on $C$ after the full iteration of (2).
\end{enumerate}
After the iteration of cycles are aborted, the algorithm returns all symmetric polynomials stored in the set $B$.

\begin{exmp}
Let us re-visit the example in Equation \eqref{eqn-ex-qg} to see how \textbf{MaxSymm}$(p,3)$ works:
\begin{align*}
p &=  2 - x_0 - x_1 - 2x_2 - 2x_3 - x_4 - x_5 + x_0x_1 + 2x_0x_2 + x_0x_3 \nonumber\\
&\qquad + x_1x_2 + 2x_1x_3 + 2x_2x_3 + 2x_2x_4 + x_2x_5 + x_3x_4 + 2x_3x_5 \nonumber\\
&\qquad - 2x_0x_1x_2 - 2x_0x_1x_3 - 2x_0x_2x_3 - 2x_1x_2x_3 \\
&\qquad - 2x_2x_3x_4 - 2x_2x_3x_5 - 2x_2x_4x_5 - 2x_3x_4x_5 + 4x_0x_1x_2x_3 + 4x_2x_3x_4x_5 \nonumber
\end{align*}
At the beginning of the algorithm, the sets $A,B$ are initialized as below (variables are represented by their indices):
$$A = \{\{0,1,2\}, \{0,1,3\}, \{0,2,3\}, \{1,2,3\}, \{2,3,4\}, \{2,3,5\}, \{2,4,5\}, \{3,4,5\} \}.$$
At the first cycle, the algorithm searches all pairs of elements in the set $B$ and $A$.
The pair $(\{0,1,2\}, \{0,1,3\})$ produces a symmetric $3$-polynomial $s=\{0,1,2,3\}$ (Line \ref{alg-symm-mp}).
Since $P_4^{(3)}(x_0,x_1,x_2,x_3$ lies in $p$, we update the set $C$ as $C = \{\{0,1,2,3\}\}$ (Line \ref{alg-b-c-c-a})
There is no further update to the set $C$ until the end of this cycle.
Since $C\neq\emptyset$, the algorithm goes into the second cycle, which starts with $B=\{\{0,1,2,3\}\}$ and $C=\emptyset$.
Since there is no other symmetric polynomial in $p$ that consists of variables more than $x_0,x_1,x_2,x_3$, there is no update to the set $C$.
The algorithm returns $\{\{0,1,2,3\}\}$ and terminates.
\end{exmp}

The algorithm \textbf{SymmRed} takes a single argument, the polynomial $p$ that requires a degree reduction. 
It uses the algorithm \textbf{MaxSymm} to find any maximal symmetric $m$-polynomials iteratively for $m$ from $3$ to $\deg p$.
For each maximal symmetric polynomial found, namely $s$, it determines a constant $a$ so that the $m$-monomials in $p$ are maximally removed by $p-a\cdot s$.
For this matter, this algorithm works properly only when the polynomial $p$ has a discrete spectrum of coefficients. 
Then the symmetric $m$-polynomial $s$ is reduced to quadratic polynomial by Theorems \ref{thm-main-1}, \ref{thm-main-2}. 
When the process of searching and reducing maximal symmetric polynomials are done, the algorithm applies the monomial reduction methods to the remaining monomials.

\begin{algorithm}
\begin{algorithmic}[1]
\Procedure{SymmRed}{$p$}\Comment{$p$ is a polynomial}
\State $q\gets0$
\For{$m\gets 3$ to $\deg p$}
	\While{$s\gets$\textbf{MaxSymm}$(p, m)$; $s\neq0$}
		\State $a\gets$ the most common coefficient of monomials in $p$ and $s$\label{alg:def-a}
		\State $p \gets p - a\cdot s$ \label{alg:p-sub}
		\If{$a>0$}
			\State $r\gets$ apply Theorem \ref{thm-main-1} to $s$ \label{alg:symm-ap}
		\Else
			\State $r\gets$ apply Theorem \ref{thm-main-2} to $-s$\label{alg:symm-an}
		\EndIf
		\State $q \gets q + |a|\cdot r$
	\EndWhile
\EndFor
\State $q \gets q+ $\textbf{MonoRed}(p) \Comment{do the monomial reduction}
\EndProcedure
\end{algorithmic}
\caption{The Symmetric Reduction algorithm}
\label{tab:symm-red-alg}
\end{algorithm}

Algorithm \ref{tab:symm-red-alg} shows a pseudocode for \textbf{SymmRed}.
The constant $a$ (Line \ref{alg:def-a} in Algorithm \ref{tab:symm-red-alg}) is determined by counting the number of occurrences of all coefficients of monomials in $p$ that lies in $s$.
However, the choice of $a$ may not be unique.
For example, suppose that $s = \{0,1,2,3\}$, $m=3$, and the polynomial $p$ is
\begin{align}
    	&-x_0x_1x_2 - x_0x_1x_3 + x_0x_2x_3 + x_1x_2x_3  \nonumber \\
    	&=\underbrace{x_0x_1x_2 + x_0x_1x_3 + x_0x_2x_3 + x_1x_2x_3}_{\textrm{Theorem } \ref{thm-main-1}}
     \underbrace{-2x_0x_1x_2 - 2x_0x_1x_3 }_{\textrm{monomial reduction}}\label{eqn:exm-symm-mono-1}\\
	&=\underbrace{-x_0x_1x_2 - x_0x_1x_3 - x_0x_2x_3 - x_1x_2x_3}_{\textrm{Theorem } \ref{thm-main-2}} 
    + \underbrace{2x_0x_2x_3 + 2x_1x_2x_3}_{\textrm{monomial reduction}}\label{eqn:exm-symm-mono-2}
\end{align}
Since there are the same numbers of $+1$ and $-1$ coefficients, we can take either Equation \eqref{eqn:exm-symm-mono-1} or \eqref{eqn:exm-symm-mono-2}.
In the actual implementation of the algorithm, we took a random choice of $a$ between candidates when this kind of situation happened.

Now, we present the efficiency of the symmetric reduction method.
We tested \textbf{SymmRed} on the objective function $Q_G$ in Equation \eqref{eqn:utility-polynomial} for random $p$-graphs $G$, where $p$ is the probability of the existence of an edge between each pair of two vertices in $G$.
We varied the number $v$ of vertices $v=3,4,5,6,7,8$ and $p=0.75$, $0.80$, $0.85$, $0.90$, $0.95$.
The number of colors are set as $v$ and thus $\lfloor\log_2(v-1)\rfloor+1$ digits are used for the binary representation of colors.
We generated $10^4$ random graphs and and their objective functions for each $p$ and $v$.
We applied the symmetric and the monomial reduction methods, and counted the numbers of total variables and monomials in the reduced quadratic polynomials.

\begin{longtable}{|c|c|c|c|c|c|c|c|c|}
    \hline
    \multirow{2}{*}{$p$} & \multirow{2}{*}{$v$} & \multirow{2}{*}{n} &
    \multicolumn{2}{c|}{\centering Symmetric reduction} & 
    \multicolumn{2}{c|}{\centering Monomial reduction} &
    \multirow{2}{*}{$\displaystyle\frac{r_1-n}{r_2-n}$} & \multirow{2}{*}{$N_1/N_2$}\\
    \cline{4-7}
    & & & $r_1$ & $N_1$ & $r_2$ & $N_2$ & &\\

\hline
\multirow{6}{*}{0.75} 
& 3 & 6 & 10.31 & 41.72 & 17.20 & 66.97 & 40.00\% & 62.29\% \\ 
& 4 & 8 & 16.90 & 75.88 & 30.41 & 125.41 & 40.00\% & 60.50\% \\ 
& 5 & 15 & 145.12 & 795.16 & 327.76 & 1597.50 & 41.61\% & 49.78\% \\ 
& 6 & 18 & 212.62 & 1182.61 & 486.06 & 2382.62 & 41.58\% & 49.63\% \\ 
& 7 & 21 & 291.53 & 1636.67 & 673.41 & 3313.83 & 41.47\% & 49.39\% \\ 
& 8 & 24 & 384.38 & 2174.45 & 894.23 & 4413.20 & 41.41\% & 49.27\% \\ 
\hline
\multirow{6}{*}{0.80} 
& 3 & 6 & 10.63 & 43.49 & 17.87 & 70.04 & 40.00\% & 62.10\% \\ 
& 4 & 8 & 17.52 & 79.96 & 31.89 & 132.63 & 40.00\% & 60.28\% \\ 
& 5 & 15 & 153.63 & 845.02 & 348.57 & 1701.93 & 41.56\% & 49.65\% \\ 
& 6 & 18 & 224.47 & 1253.53 & 515.08 & 2528.39 & 41.54\% & 49.58\% \\ 
& 7 & 21 & 309.64 & 1743.23 & 717.00 & 3532.79 & 41.47\% & 49.34\% \\ 
& 8 & 24 & 407.97 & 2315.10 & 951.84 & 4702.65 & 41.38\% & 49.23\% \\ 
\hline
\multirow{6}{*}{0.85} 
& 3 & 6 & 10.98 & 45.53 & 18.64 & 73.59 & 40.00\% & 61.87\% \\ 
& 4 & 8 & 18.20 & 84.52 & 33.53 & 140.75 & 40.00\% & 60.05\% \\ 
& 5 & 15 & 162.28 & 895.16 & 369.25 & 1805.79 & 41.58\% & 49.57\% \\ 
& 6 & 18 & 237.42 & 1331.70 & 546.97 & 2688.59 & 41.48\% & 49.53\% \\ 
& 7 & 21 & 326.47 & 1841.98 & 758.43 & 3740.97 & 41.42\% & 49.24\% \\ 
& 8 & 24 & 431.16 & 2453.30 & 1007.69 & 4983.24 & 41.39\% & 49.23\% \\ 
\hline
\multirow{6}{*}{0.90} 
& 3 & 6 & 11.35 & 47.78 & 19.47 & 77.54 & 40.00\% & 61.62\% \\ 
& 4 & 8 & 18.82 & 88.81 & 35.07 & 148.39 & 40.00\% & 59.85\% \\ 
& 5 & 15 & 170.38 & 942.11 & 389.19 & 1905.94 & 41.52\% & 49.43\% \\ 
& 6 & 18 & 249.12 & 1404.09 & 577.36 & 2841.28 & 41.32\% & 49.42\% \\ 
& 7 & 21 & 344.49 & 1946.97 & 803.22 & 3966.03 & 41.36\% & 49.09\% \\ 
& 8 & 24 & 454.33 & 2594.31 & 1065.11 & 5271.75 & 41.33\% & 49.21\% \\ 
\hline
\multirow{6}{*}{0.95} 
& 3 & 6 & 11.69 & 49.87 & 20.23 & 81.22 & 40.00\% & 61.41\% \\ 
& 4 & 8 & 19.40 & 92.83 & 36.51 & 155.55 & 40.00\% & 59.68\% \\ 
& 5 & 15 & 178.64 & 989.51 & 409.39 & 2007.44 & 41.49\% & 49.29\% \\ 
& 6 & 18 & 259.17 & 1470.89 & 608.56 & 2998.07 & 40.84\% & 49.06\% \\ 
& 7 & 21 & 360.95 & 2041.47 & 844.74 & 4174.64 & 41.27\% & 48.90\% \\ 
& 8 & 24 & 474.11 & 2722.52 & 1121.66 & 5555.88 & 41.01\% & 49.00\% \\ 
\hline
\multirow{6}{*}{1.00}
& 3 & 6 & 12 & 52 & 21 & 85 & 40.00\% & 61.18\% \\ 
& 4 & 8 & 20 & 97 & 38 & 163 & 40.00\% & 59.51\% \\ 
& 5 & 15 & 187 & 1037 & 430 & 2111 & 41.45\% & 49.12\% \\ 
& 6 & 18 & 264 & 1516 & 639 & 3151 & 39.61\% & 48.11\% \\ 
& 7 & 21 & 378 & 2137 & 889 & 4397 & 41.13\% & 48.60\% \\ 
& 8 & 24 & 480 & 2797 & 1180 & 5849 & 39.45\% & 47.82\% \\ 
\hline
   \caption{The average numbers of total variables $r_i$ and monomials $N_i$ in the quadratic polynomials reduced from the objective function of the graph coloring problem on the random $p$-graphs, obtained by the symmetric reduction method ($i=1$) and the monomial reduction methods ($i=2$). The $v$ is the number of vertices in the random graph and $n$ is the number of variables in the objective function. The last two columns are ratios between averages on numbers of auxiliary variables ($(r_1-n)/(r_2-n)$) and the total numbers of monomials $N_1/N_2$.}\label{tab:avg-num}\\
\end{longtable}

Table \ref{tab:avg-num} shows the summarized results.
The $r_1$ and $r_2$ on the fourth and sixth columns of the table are the average numbers of all variables in the reduced quadratic polynomials obtained by applying the symmetric and monomial reduction methods respectively.
The $N_1$ and $N_2$ on the fifth and seventh column of the table are the average numbers of all monomials in the reduced quadratic polynomials obtained by the symmetric and monomial reduction methods respectively.
We take the ratio between average numbers of auxiliary variables in the reduced quadratic polynomial obtained by the symmetric and monomial reduction methods.
The last rows of Table \ref{tab:avg-num} shows the cases for the complete graphs.

\begin{figure}[!b]
\centering
    \begin{subfigure}[b]{0.3\textwidth}
    \centering
    \includegraphics[scale=.4]{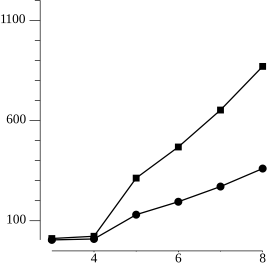}
    \caption{$p=0.75$}
    \end{subfigure}
    \begin{subfigure}[b]{0.3\textwidth}
    \centering
    \includegraphics[scale=.4]{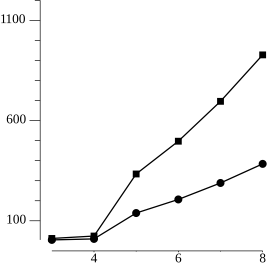}
    \caption{$p=0.80$}
    \end{subfigure}
    \begin{subfigure}[b]{0.3\textwidth}
    \centering
    \includegraphics[scale=.4]{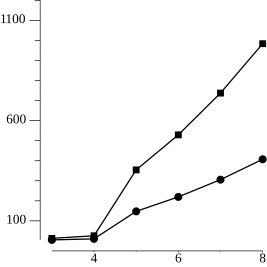}
    \caption{$p=0.85$}
    \end{subfigure}
\end{figure}

\begin{figure}
\ContinuedFloat
\centering
    \begin{subfigure}[b]{0.3\textwidth}
    \centering
    \includegraphics[scale=.4]{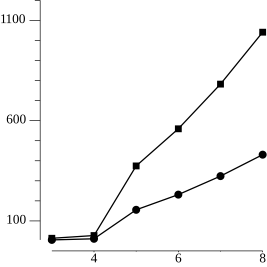}
    \caption{$p=0.90$}
    \end{subfigure}
    \begin{subfigure}[b]{0.3\textwidth}
    \centering
    \includegraphics[scale=.4]{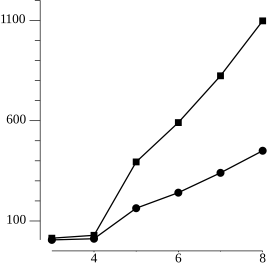}
    \caption{$p=0.95$}
    \end{subfigure}
    \begin{subfigure}[b]{0.3\textwidth}
    \centering
    \includegraphics[scale=.4]{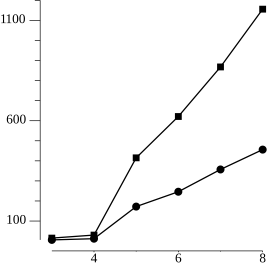}
    \caption{$p=1.00$}
    \end{subfigure}
    \caption{Plots on the average numbers of auxiliary variables in the quadratic polynomial reduced from the objective functions of the graph coloring problems on random $p$-graphs by the symmetric reduction method (circle-dot) and the monomial reductions (square-dots). The number of colors are set to be the number of vertices of the graph. The horizontal axis represents the number of vertices and the vertical axis represents the average numbers of auxiliary variable.}
    \label{fig:growth}
\end{figure}

Figure \ref{fig:growth} shows the data points for $(v, r_1)$ and $(v, r_2)$ for each $p$.
The symmetric reduction method produces only $39.45\%\sim 41.61\%$ of auxiliary variables than the monomial reduction method.
The average number of monomials in the reduced polynomial obtained by the symmetric reduction method are approximately $47.82\%\sim 62.29\%$ of what produced by the monomial reduction method.

\section{Conclusion}

One of the main limitation that the current quantum annealing systems have, in solving real-world problems, is the number of qubits of the quantum annealing machines.
Thus minimizing the number of variables in a QUBO formulation is essential.
We showed that the symmetric reduction method can produces substantially less number of auxiliary variables than the conventional methods for the QUBO formulation of the graph coloring problem.

\vskip 6mm
\bibliographystyle{amsplain}

\end{document}